\documentclass{Interspeech}

\interspeechcameraready

\title{Smooth Operators: LLMs Translating Imperfect Hints into Disfluency-Rich Transcripts}

\author[affiliation={1}]{Duygu}{Altinok}

\affiliation{}{Independent Researcher}{Germany}
\email{duygu.altinok@onlyduygu.com}

\usepackage{comment}
\usepackage[most]{tcolorbox}
\usepackage{booktabs} 
\usepackage{multirow} 
\usepackage{adjustbox}
\usepackage{xurl}

\begin{document}

\maketitle

\begin{abstract}
Accurate detection of disfluencies in spoken language is crucial for enhancing the performance of automatic speech and language processing systems, as well as fostering the development of more inclusive speech and language technologies. Leveraging the growing trend of large language models (LLMs) as versatile learners capable of processing both lexical and non-lexical inputs (e.g., audio and video), we propose a novel approach to transcribing disfluencies as explicit tokens with timestamps, enabling the generation of fully annotated disfluency-rich transcripts. Our method integrates acoustic representations extracted from an audio encoder with textual inputs of varying quality: clean transcriptions without disfluencies, time-aligned transcriptions from aligners, or outputs from phoneme-based ASR models—all of which may contain imperfections. Importantly, our experiments demonstrate that textual inputs do not need to be flawless. As long as they include timestamp-related cues, LLMs can effectively smooth the input and produce fully disfluency-annotated transcripts, underscoring their robustness in handling imperfect hints.
\end{abstract}

\section{Introduction}
Recent advancements in automatic speech recognition (ASR) \cite{bhandari2025reverbopensourceasrdiarization, radford2022robustspeechrecognitionlargescale, puvvada2024moreaccuratespeechrecognition} have achieved human parity in high-resource languages. However, speech involves more than words—it includes phonemes, prosody, and non-word cues like disfluencies, which are crucial for understanding intent. Disfluencies, such as repetitions, hesitations, and replacements \cite{lian2023unconstraineddysfluencymodelingdysfluent}, are essential for dialogue systems and clinical applications like speech therapy. Yet, research on disfluency detection is limited by dataset constraints, sparse annotations, and a narrow focus on stutter detection, often treated as a classification task. Identifying both the type and timing of disfluencies remains a challenge \cite{lian2023unconstraineddysfluencymodelingdysfluent}.

Early approaches to disfluency modeling treated it as a time-based object detection problem, focusing on temporal location and disfluency types \cite{lian2023unconstraineddysfluencymodelingdysfluent, lian-anumanchipalli-2024-towards}. Token-based methods, such as TimeTokens \cite{zhou2024timetokensbenchmarkingendtoend}, have since emerged, addressing disfluency detection as a transcription task that captures both type and timing in a unified framework. Whisper-based tokenization \cite{radford2022robustspeechrecognitionlargescale} further demonstrated the feasibility of combining transcription and timestamp generation, paving the way for more robust disfluency modeling.

\begin{figure}[ht]
  \centering
  \includegraphics[scale=0.18]{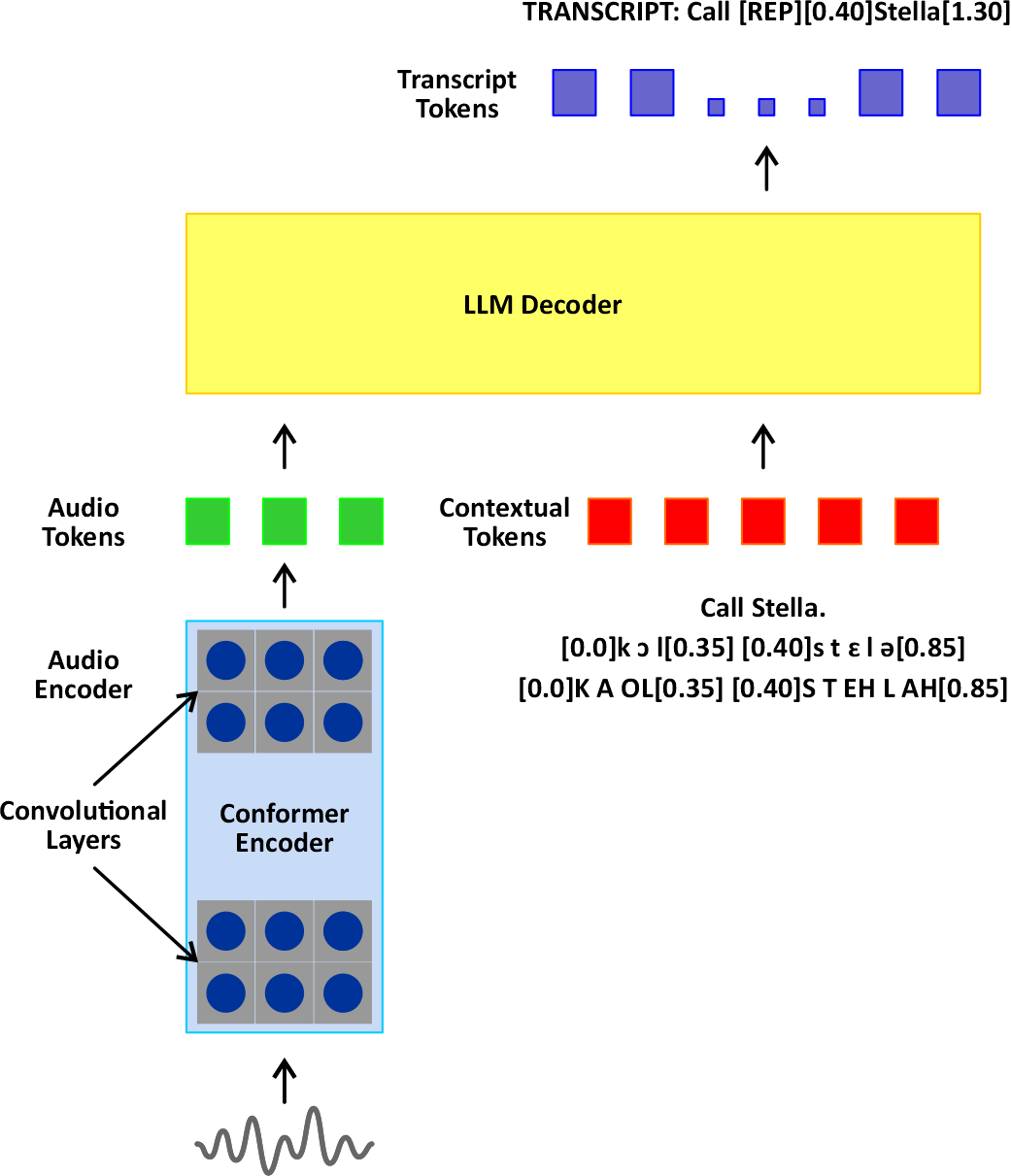} 
  \caption{Overall model architecture.}
  \label{fig:model} 
\end{figure}

Recent research has also explored integrating large language models (LLMs) \cite{openai2024gpt4technicalreport, touvron2023llamaopenefficientfoundation} into disfluency detection. For instance, \cite{wagner2024largelanguagemodelsdysfluency} combined audio embeddings from Wav2Vec \cite{baevski2020wav2vec20frameworkselfsupervised} with ASR hypotheses generated by Whisper, while CrisperWhisper \cite{wagner2024crisperwhisperaccuratetimestampsverbatim} improved transcription robustness. However, current methods often rely on imperfect text inputs or focus on specific disfluency types, limiting their generalizability.

In this work, we propose Smooth-LLaMa, a novel framework leveraging LLMs to transcribe disfluency tokens and generate accurate timestamps by combining audio embeddings and textual inputs. Our model employs a Conformer-based \cite{gulati2020conformerconvolutionaugmentedtransformerspeech} acoustic encoder and integrates phoneme- and word-level alignments from Wav2VecPhoneme \cite{xu2021simpleeffectivezeroshotcrosslingual} and MFA (Montreal Forced Aligner) aligners \cite{mcauliffe2017montreal}. Despite imperfections in textual inputs, Smooth-LLaMa produces highly accurate disfluency-annotated transcripts by smoothing and integrating the available information. During training, textual inputs serve as soft guidance without contributing to the cross-entropy loss, allowing the framework to operate flexibly with audio-only inputs during inference. Smooth-LLaMa also improves imperfect transcriptions, generating verbatim annotations and enriching disfluency datasets.

Our framework unifies token- and time-based approaches by generating disfluency tokens and timestamps simultaneously, capturing both contextual and temporal information. Evaluations on the VCTK-TTS dataset \cite{zhou24e_interspeech} demonstrate significant improvements in transcription quality, disfluency detection, and timestamp accuracy, establishing Smooth-LLaMa as a state-of-the-art solution for disfluency modeling.
Our contributions are summarized as follows:
\begin{itemize}
\item We propose a robust framework, Smooth-LLaMa, that leverages LLMs for disfluency transcription and timestamp generation using audio and textual inputs.
\item We demonstrate that imperfect textual inputs can still guide LLMs for flexible and robust disfluency annotation.
\item We unify token- and time-based approaches, achieving state-of-the-art results on the VCTK-TTS dataset.
\item To our knowledge, this is the first study to combine LLMs with disfluency transcription, generating both type and duration annotations.
\end{itemize}

Our results show that LLMs, when prompted correctly, can effectively transcribe disfluencies with type and duration annotations, improving ASR performance. To our knowledge, this is the first study to use LLMs for comprehensive disfluency transcription.

\section{Dataset}
In our work, we used the VCTK-TTS \cite{zhou24e_interspeech} dataset, which was created using a pipeline that simulates disfluencies in the text domain. Base sentences were converted into IPA phoneme sequences, and TTS rules were applied to edit phonemes and simulate disfluencies. The dataset includes annotations for both word-level and phoneme-level disfluencies. A summary of the dataset statistics is provided in Table \ref{tab:vctk-stats}.

\begin{table}[h!]
\centering
\caption{Statistics of VCTK-TTS dataset}
\label{tab:vctk-stats}
\begin{tabular}{|l|l|}
\hline
\textbf{Dysfluency}       & \textbf{Hours} \\ \hline
Repetition                &   114.92            \\ \hline
Missing/Deletion          &   122.05             \\ \hline
Block/Pause               &   68.56              \\ \hline
Replace/Substitution      &   64.33               \\ \hline
Prolongation              &   57.06               \\ \hline
Insertion                 &    -                \\ \hline
\textbf{Total}      &     426.93          \\ \hline
\end{tabular}
\end{table}

Each instance in the dataset consists of a clean transcript (e.g., "Call Stella"), audio files, and corresponding annotations in JSON format. Word-level annotations include words, timestamps, and disfluency types, e.g., [(Call, 0.00, 0.05, none), (Stella, 0.08, 0.20, PROLONG)]. Phoneme-level annotations detail the phoneme affected by the disfluency, e.g., [(EH, 0.10, 0.15, PROLONG)].

For data preparation, we processed word-level and phoneme-level disfluencies separately. Figure \ref{fig:data-making} illustrates the process used for preparing textual inputs.

\section{Method}
\subsection{Preparing Textual Data}

\textbf{Preparing the Textual Input:} To prompt the LLM effectively, we prepared various types of textual input. Below are the formats used in our experiments:

\begin{itemize}
\item \textbf{wav2vec-phonemes:} To simulate scenarios without a clean transcript, we fed the audio data into a Wav2VecPhoneme model trained with CTC. This model outputs IPA phonemes with timestamps. Since the dataset uses ARPAbet annotations, the IPA phonemes were converted to ARPAbet.
\item \textbf{wav2vec-words:} Using the phoneme outputs from the Wav2Vec-Phoneme model, we generated word-level sequences with timestamps. This mapping was achieved using rule-based heuristics to align phonemes to words.
\item \textbf{clean-transcript:} The clean transcript provided by the dataset was used directly. Note that this transcript does not align perfectly with the audio.
\item \textbf{aligned-phonemes:} Using the clean transcript and audio, we generated phoneme-level alignments with timestamps using the MFA aligner. The resulting text is a sequence of phonemes, each accompanied by start and end times.
\item \textbf{aligned-words:} Similarly, we used the MFA aligner to generate word-level alignments with timestamps.
\end{itemize}

\begin{figure*}[ht]
  \centering
  \includegraphics[width=\linewidth]{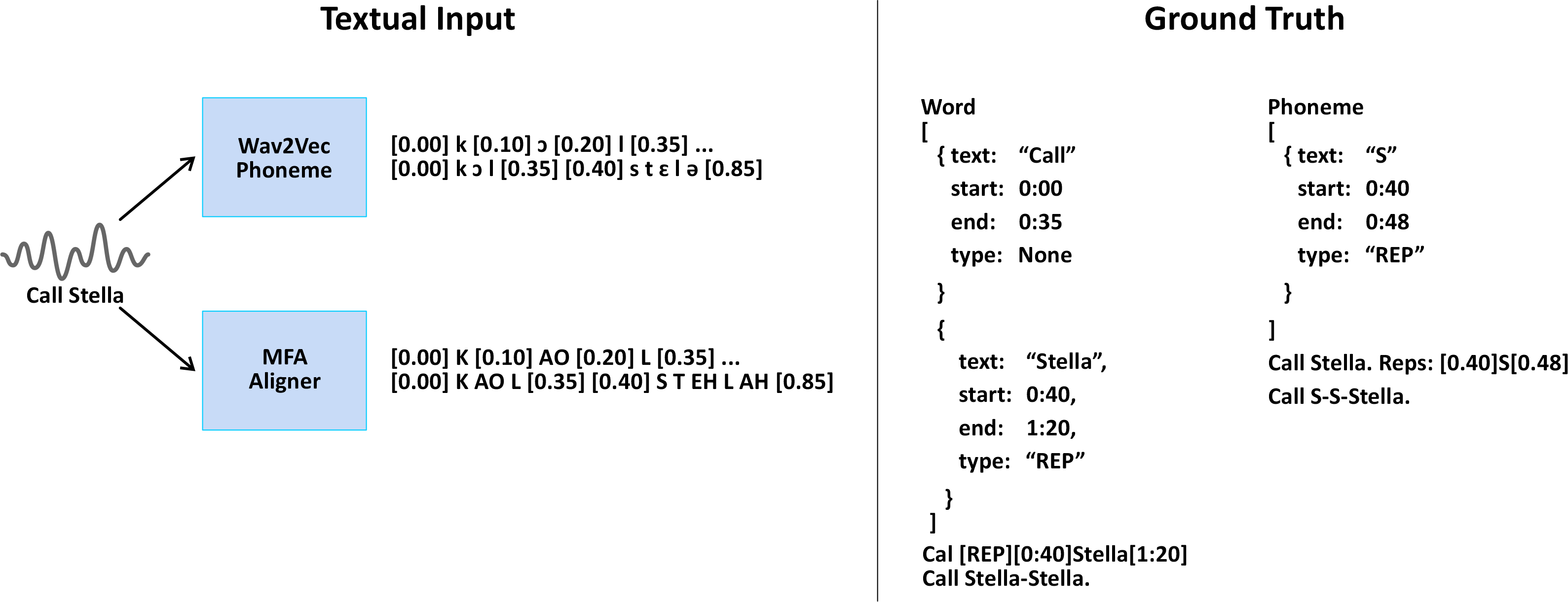} 
  \caption{Different types of textual data preparation are on the left, and the ground truth preparation process is on the right.}
  \label{fig:data-making} 
\end{figure*}

Figure \ref{fig:data-making} illustrates examples of each input type. In our experiments, we tested each type individually as well as all types combined. For the combined input, the prompts were concatenated directly without any delimiters.

\noindent \textbf{Ground Truth:} Ground truth annotations were derived from the dataset JSON files. For word-level disfluencies, we inserted disfluency tokens at appropriate positions:[BLOCK] for long pauses, [REP] for repetitions, [MISS] for omissions, [PRO] for prolongations.
Each token was surrounded by timestamps, but only for disfluent segments. For phoneme-level disfluencies, the annotations consisted of the clean transcript concatenated with a list of phonemes, where each phoneme was accompanied by its timestamps and disfluency type.
Additionally, the ground truth transcripts were prepended with the prompt \textbf{TRANSCRIPT:}. Figure \ref{fig:data-making} illustrates examples of the ground truth texts.

\subsection{Model}

Figure \ref{fig:model} illustrates the overall architecture of our model.

\noindent\textbf{LLM:}  
We used a pretrained LLaMa 3 model \cite{touvron2023llamaopenefficientfoundation} with 8B parameters as the decoder.

\noindent\textbf{Audio Encoder:}  
The audio encoder begins with four downsampling blocks, resulting in a 16x reduction in the temporal resolution of the audio representations. This is followed by a stack of Conformer blocks \cite{gulati2020conformerconvolutionaugmentedtransformerspeech} with rotary positional embeddings \cite{su2023roformerenhancedtransformerrotary}, a hidden dimensionality of 1024, and a kernel size of 9. At the end of the encoder, an additional downsampling block and a linear layer are applied. Consequently, the decoder observes audio tokens sampled every 320ms, with each token having a dimensionality of 4096, matching the LLaMa model's embedding dimension.  

We initialized the audio encoder using a pretrained Conformer model and finetuned it on our dataset. The input features are 80-dimensional log Mel spectrograms, computed with a 25ms window and a 10ms stride.

\noindent\textbf{Textual Input:}  
Textual input is tokenized using the LLaMa tokenizer to produce textual tokens. These textual tokens are concatenated with the audio tokens to form the input to the decoder. Note that positional embeddings are not explicitly applied at this stage, as the rotary positional embeddings are already integrated into the attention layers of the Conformer blocks.

\noindent\textbf{Objective Function:}  
During training, the cross-entropy loss is calculated only on the audio tokens, while masking the textual tokens. Let $\mathbf{p} = \{p_1, p_2, \dots, p_M\}$ denote the textual tokens, and $\mathbf{t} = \{t_1, t_2, \dots, t_N\}$ the transcript tokens. The training objective is defined as:

\begin{equation}
    \mathcal{L} = -\sum_{i=M+1}^{M+N} \log P(t_i \mid \mathbf{p}, \mathbf{t}_{<i}, \mathbf{z}_{\text{enc}})
\end{equation}

where $\mathbf{z}_{\text{enc}} \in \mathbb{R}^{N \times d_{\text{embed}}}$ represents the audio embeddings, and $d_{\text{embed}}$ is the embedding dimension of the LLaMa model.

\subsection{Evaluation Metrics}
We adopted metrics from \cite{zhou2024timetokensbenchmarkingendtoend} to enable direct comparison with prior work. These metrics evaluate both timing accuracy and disfluency type detection.  

\textbf{Token Error Rate (TER)} measures transcription accuracy by calculating the percentage of substitutions, deletions, and insertions compared to the reference text. \textbf{Disfluency Existence Accuracy (EAcc.)} evaluates the correct detection of disfluencies in speech utterances, while \textbf{Disfluency Classification Accuracy (CAcc.)} measures the accuracy of identifying specific disfluency types.  

\textbf{Bound Loss (BL)} \cite{redmon2016lookonceunifiedrealtime} computes the mean squared error between predicted and actual disfluent region boundaries, sampled at 20ms intervals. \textbf{Token Distance (TD)} measures the token-level displacement between predicted and actual disfluency positions in the text.

\subsection{Experimental Setup}
In our experiments, we first fine-tuned the Conformer model while keeping the rest of the network, including LLaMa, frozen. Subsequently, we performed joint training.

\noindent\textbf{Textual Input:} We experimented with five types of input text: wav2vec-phonemes, wav2vec-words, clean-transcript, aligned-phonemes, and aligned-words, concatenated with whitespace.

\noindent\textbf{Finetuning Conformer:} We used the Wav2Vec 2.0 Conformer Large model (370M parameters) with RoPE embeddings\footnote{\url{https://huggingface.co/facebook/wav2vec2-conformer-rope-large-960h-ft}}, pretrained on 960 hours of LibriSpeech data \cite{7178964}. Finetuning was performed using the HuggingFace Trainer \cite{wolf2020huggingfacestransformersstateoftheartnatural} with a batch size of 32, a learning rate of $2e-5$, and 1250 warmup steps. Training lasted 10 epochs, with AdamW optimization (weight decay = 0.01). The latent feature extractor was frozen for the first 2 epochs to avoid catastrophic forgetting, then unfrozen for the remaining epochs.

\noindent\textbf{Training:} To optimize performance and memory usage, we used parameter-efficient techniques and quantization. Using BitsandBytes\footnote{\url{https://huggingface.co/docs/transformers/main/quantization/bitsandbytes}}, we applied 4-bit weight quantization in the NF4 format \cite{dettmers2023qloraefficientfinetuningquantized}, performing computations in bfloat16 \cite{bfloat16_google_cloud} for stability and efficiency.

We applied Low-Rank Adaptation (LoRA) \cite{hu2021loralowrankadaptationlarge} using the HuggingFace PEFT library\footnote{\url{https://huggingface.co/docs/peft/index}}, with rank-16 matrices, a scaling factor of 32, and a dropout rate of 0.01, focusing on attention layers. Training was conducted for 100 epochs using AdamW \cite{loshchilov2019decoupledweightdecayregularization} (learning rate = $2e-4$, weight decay = 0.01). All epochs were necessary, as LLaMa struggled to generate timestamps early on.

\noindent\textbf{Inference:} During training, both textual and audio tokens were used; however, all inferences were performed using audio tokens only.

All experiments were performed on two NVIDIA H100 GPUs. Our code is publicly available on GitHub\footnote{\url{https://github.com/DuyguA/Interspeech2025-Smooth-Operating-LLMs-for-Disfluency}}.

\section{Results and Discussion}
We evaluated our approach using the dataset from \cite{zhou2024timetokensbenchmarkingendtoend}, enabling direct comparison with their Whisper-based approach, referred to as the Whisper Detector. Table \ref{tab:comparison_metrics} summarizes the performance of our model against this baseline.

Our approach outperforms the Whisper-based method across all metrics, with particularly significant improvements in the Token Distance (TD) metric, highlighting the accuracy of the resulting transcripts. This improvement may be attributed to the broader language understanding capabilities of LLaMa compared to Whisper, which is primarily designed as an ASR model. Additionally, while Whisper is prone to hallucinations, we did not observe any hallucinations in the transcripts generated by our model on the test set, despite LLMs being known for such behavior.

Furthermore, the previous benchmark \cite{zhou2024timetokensbenchmarkingendtoend} reported an average Bound Loss (BL) of 23ms. In contrast, our model achieved an average BL of 12ms, demonstrating superior performance in capturing timing accuracy.

\begin{table*}[!ht]
\caption{Comparison of performance metrics between Whisper Detector - abbreviated as WhisperD and our proposed method. Token Distance is not applicable to our work at the phoneme level because we only output disfluent phonemes, not the entire transcript.}
\label{tab:comparison_metrics}
\centering
\begin{tabular}{llcccccccc}
\toprule
\textbf{Levels} & \textbf{Types} & \multicolumn{2}{c}{\textbf{TER (\%, ↓)}} & \multicolumn{2}{c}{\textbf{EAcc. (\%, ↑)}} & \multicolumn{2}{c}{\textbf{CAcc. (\%, ↑)}} & \multicolumn{2}{c}{\textbf{TD (e-3, ↓)}} \\
\cmidrule(lr){3-4} \cmidrule(lr){5-6} \cmidrule(lr){7-8} \cmidrule(lr){9-10}
 &  & \textbf{WhisperD} & \textbf{Ours} & \textbf{WhisperD.} & \textbf{Ours} & \textbf{WhisperD.} & \textbf{Ours} & \textbf{WhisperD.} & \textbf{Ours} \\
\midrule
\multirow{4}{*}{Word} 
    & Repetition   & 0.144 & 0.05 & 99.57 & 99.95 & 99.36 & 99.45    & 0.91 & 0.43    \\
    & Deletion     & 0.283 & 0.07 & 97.98 & 99.92 & 96.85 & 99.96    & 1.01 & 0.52   \\
    & Insertion    & 0.212 & 0.08 & 98.93 & 99.91 & 97.47 & 99.33    & 3.03 & 1.45    \\
    & Pause        & 0.195 & 0.04 & 99.04 & 99.95 & 98.28 & 99.41    & 7.00 & 2.65    \\
\midrule
\multirow{4}{*}{Phoneme} 
    & Repetition & 0.154 & 0.03 & 98.95 & 99.91 & 99.23 & 99.96 & 1.87 & - \\
    & Deletion  & 1.141 & 0.05 & 98.84 & 99.12 & 98.17 & 99.11 & 6.46 & - \\
    & Substitution & 0.335 & 0.03 & 96.11 & 99.41 & 95.75 & 99.02 & 7.82 & - \\
    & Prolongation & 0.326 & 0.08 & 98.89 & 99.93 & 98.81 & 99.84 & 2.43 & - \\
\bottomrule
\end{tabular}
\end{table*}

\subsection{Ablation Studies}
\subsubsection{Effect of Textual Input}
In the experiments we concatenated all the textual input sorts that we created and fed as a single input. Here, we experimented with different combinations of textual inputs to understand the effect of the textual input.

\begin{table}[h!]
\caption{Word-level performance metrics for different input types, including TER, EAcc, CAcc, TD, and BL, averaged across the entire dataset.}
\label{tab:world-comp}
\centering
\begin{tabular}{@{}lccccc@{}}
\toprule
\textbf{Input Type}       & \textbf{TER} & \textbf{EAcc.} & \textbf{CAcc.} & \textbf{TD} & \textbf{BL} \\ \midrule
All  & 0.06   & 99.93 & 99.53  & 1.26 & 12 \\
wav2vec-word & 0.06  & 99.93 & 99.53 & 1.26  & 15 \\
wav2vec-phon & 0.06 & 99.93  & 99.53 & 1.26 & 12  \\ 
aligned-word & 0.06  & 99.93 & 99.53 & 1.26  & 15 \\
aligned-phon & 0.06 & 99.93  & 99.53 & 1.26 & 12  \\ 
clean-trans & 0.15 & 95.00 & 92.00 & 3.50 & 25 \\
None & 0.45 & 85.00 & 80.00 & 7.00 & 45 \\
\bottomrule
\end{tabular}
\end{table}

Our results show that using a single type of textual input with timing information is sufficient, removing the need for multiple inputs. As shown in Table \ref{tab:world-comp}, word-level timestamps provide strong semantic guidance for predicting tokens accurately in fluent speech but struggle with fine-grained alignment in disfluent utterances like repetitions or elongations (e.g., "st-st-stella"). In such cases, disfluencies are assigned to broad word-level intervals, leading to imprecise timestamps despite correct token predictions. This makes word-level timestamps suitable for coarse-grained tasks but inadequate for capturing disfluency nuances.

\begin{table}[ht]
\caption{Phoneme-level performance metrics for different input types, including TER, EAcc, CAcc, TD, and BL, averaged across the entire dataset.}
\label{tab:phoneme_comparison}
\centering
\begin{tabular}{@{}lccccc@{}}
\toprule
\textbf{Input Type} & \textbf{TER} & \textbf{EAcc.} & \textbf{CAcc.} & \textbf{TD} & \textbf{BL} \\ \midrule
All  & 0.04   & 99.59 & 99.48  & - & 10 \\
wav2vec-word & 0.12  & 97.00 & 94.00 & -  & 18 \\
wav2vec-phon & 0.04   & 99.59 & 99.48  & - & 10  \\ 
aligned-word &0.12  & 97.00 & 94.00 & -  & 18 \\
aligned-phon & 0.04   & 99.59 & 99.48  & - & 10  \\ 
clean-trans & 0.25 & 90.00 & 85.00 & - & 35 \\
None & 0.60 & 75.00 & 70.00 & - & 60 \\\bottomrule
\end{tabular}
\end{table}

Phoneme-level timestamps, however, align audio embeddings to sub-word features, enabling precise token and timestamp predictions even in complex disfluent speech, as shown in Table \ref{tab:phoneme_comparison}. By offering fine-grained temporal supervision, phoneme-level inputs excel in capturing disfluencies like repetitions and elongations. Without them, Bound Loss (BL) increases, reducing alignment precision. While word-level timestamps work for simpler tasks, phoneme-level timestamps are crucial for detailed disfluency annotation and accurate alignment.

Overall, results show poor performance without textual inputs (TER: 0.60, BL: 60), while clean transcripts improve token accuracy but lack precise alignment (TER: 0.25, BL: 35). Word-level timestamps reduce TER to 0.12 and BL to 18, but phoneme-level timestamps achieve the best results (TER: 0.04, BL: 9–10). Transitioning from word-level to phoneme-level timestamps reduces TER by 67\% and improves BL by 50

\subsection{Effect of LLM Size}
When scaling down from an 8B model to smaller ones (1B and 3B), both Token Error Rate (TER) and Bound Loss (BL) worsened due to reduced model capacity. For the 1B model, TER increased significantly—by 0.10–0.15 across all input types—while BL nearly doubled for phoneme-level inputs, reflecting poorer alignment precision. The 3B model showed less degradation, with TER rising by 0.05–0.08 and BL increasing by about 50\% compared to the 8B model. While the 3B model retained some phoneme-level precision, the 1B model struggled with fine-grained alignment, leading to higher errors. Aligning audio with text is challenging, and neither the 1B nor 3B models were sufficient for this task.

\section{Conclusion}
This work demonstrates how large language models (LLMs) can smooth and integrate imperfect audio and textual inputs for disfluency transcription and timestamp generation. Smooth-LLaMa leverages contextual understanding to refine noisy data, achieving precise outputs even with imperfect textual inputs, such as word- or phoneme-level alignments. The model’s ability to adapt and unify diverse inputs highlights its robustness and versatility, particularly in handling disfluent speech. By smoothing inconsistencies across modalities, LLMs prove to be powerful tools for generating accurate and high-quality transcriptions, setting a new benchmark for speech processing tasks.

\bibliographystyle{IEEEtran}
\bibliography{mybib}

\end{document}